\documentclass[twocolumn,superscriptaddress,showpacs,amsmath,amssymb]{revtex4-1}

\usepackage{graphicx}
\usepackage{dcolumn}
\usepackage{bm}
\usepackage{CJK}
\usepackage{txfonts}
\usepackage{color}
\usepackage{tabularx}
\usepackage{multirow}
\usepackage{hhline}%
\usepackage{hyperref}

\def\beq{\begin{equation}}
\def\eeq{\end{equation}}
\def\bea{\begin{eqnarray}}
\def\eea{\end{eqnarray}}

\def\fun#1#2{\lower3.6pt\vbox{\baselineskip0pt\lineskip.9pt
  \ialign{$\mathsurround=0pt#1\hfil##\hfil$\crcr#2\crcr\sim\crcr}}}

\usepackage{ulem}
\newcommand{\delete}{\bgroup\markoverwith{\textcolor{red}{\rule[0.5ex]{2pt}{1pt}}}\ULon}

\begin{document}
\title{Central depressions in the charge density profiles of the nuclei around $^{46}$Ar}

\author{Jun Ling Song }
\author{Qiang Zhao }
\author{Wen Hui Long }\email{longwh@lzu.edu.cn}
\affiliation{School of Nuclear Science and Technology, Lanzhou University, Lanzhou 730000, China}
\affiliation{Key Laboratory of Special Function Materials and Structure Design, Ministry of Education, Lanzhou 730000, China}

\begin{abstract}
  The occurrence of the proton bubble-like structure has been studied within the relativistic Hartree-Fock-Bogoliubov (RHFB) and relativistic Hartree-Bogoliubov (RHB) theories by exploring the bulk properties, the charge density profiles and single proton spectra of argon isotopes and $N = 28$ isotones. It is found that the RHFB calculations with PKA1 effective interaction, which can properly reproduce the charge radii of argon isotopes and the $Z=16$ proton shell nearby, do not support the occurrence of the proton bubble-like structure in argon isotopes due to the prediction of deeper bound proton orbit $\pi2s_{1/2}$ than $\pi1d_{3/2}$. For $N = 28$ isotones, $^{42}$Si and $^{40}$Mg are predicted by both RHFB and RHB models to have the proton bubble-like structure, owing to the large gap between the proton $\pi2s_{1/2}$ and $\pi1d_{5/2}$ orbits, namely the $Z=14$ proton shell. Therefore, $^{42}$Si is proposed as the potential candidate of proton bubble nucleus, which has longer life-time than $^{40}$Mg.
\end{abstract}


\maketitle
\section{INTRODUCTION}

With the worldwide development of the radioactive ion beam facilities since the 1980s, the study of the exotic nuclei becomes a hot frontier of nuclear physics \cite{Casten2000Prog.Part.Nucl.Phys.S171}. The exotic nuclei is a kind of nuclei with extreme neutron-proton ratio, which exhibit quite different features from the stable nuclei or the ones near stability line. As one of the representatives with the nuclear novel phenomena, the bubble nucleus is an exotic one characterized by the distinct central depressions of the matter distributions, namely the bubble-like structure. The study of bubble nuclei can be traced to the pioneering work of Wilson as early as in the 1940s \cite{Wilson1946Phys.Rev.538}, in which the nucleus was assumed to be a thin spherical shell to explain the equally spaced nuclear single-particle levels. After that in the 1970s, a variety of nuclear models, including the liquid drop model \cite{Swiatecki1983Phys.Scr.349}, the Thomas-Fermi model \cite{Saunier1974Phys.Lett.B293}, and the Hartree-Fock method \cite{Campi1973Phys.Lett.B291}, have been applied to test the existence of the nuclear bubble-like structure. Usually, it is recognized that the bubble-like structure originates from the low occupancy of $s$-orbit near the Fermi surface, which is the only wave function with non-zero value at $r = 0$.

The nuclear charge distribution is an important observable which can provide very detailed information of nuclear structure \cite{Hofstadter1956Rev.Mod.Phys.214, Forest1966Adv.Phys.1, Donnelly1975Annu.Rev.Nucl.Part.Sci.329}. For instance, the charge distribution can reflect the proton-density distribution in nucleus, and a central depressed charge distribution is the consequence of the proton bubble-like structure. Hence, a proton-bubble nucleus can be identified experimentally from the measurement of the charge distribution, e.g., by the elastic electron-nucleus scattering experiment. In the near future, more experimental data about the charge distribution of exotic nuclei are expected to be obtained both from the project of the SCRIT \cite{Suda2005Prog.Part.Nucl.Phys.417} and ELISe \cite{Antonov2011Nucl.Instrum.Methods_Phys.Res.A60}.


In medium mass region of the nuclear chart, some candidates of the proton-bubble nuclei were predicted, such as $^{34}$Si \cite{Grasso2009Phys.Rev.C034318, Nakada2013Phys.Rev.C067305, Wang2011Chin.Phys.Lett.102101} and $^{46}$Ar \cite{Todd2004Phys.Rev.C021301, Chu2010Phys.Rev.C024320, Khan2008Nucl.Phys.A37, Wang2011Chin.Phys.Lett.102101, Wang2011Phys.Rev.C044333}. For $^{46}$Ar, the occurrence of proton bubble-like structure is found to essentially depend on the order of the proton orbits $\pi2s_{1/2}$ and $\pi1d_{3/2}$, while the prediction is evidently model dependent. In the calculations of the nonrelativistic Skyrme Hartree-Fock-Bogoliubov(HFB) approach \cite{Wang2011Phys.Rev.C044333}, it is found that the proton bubble-like structure may emerge in $^{46}$Ar as the conclusion of the inversion of proton $\pi2s_{1/2}$ and $\pi1d_{3/2}$ orbits with tensor interaction included. Similarly, the relativistic mean-field theory (RMF) models with nonlinear meson self-couplings \cite{Todd2004Phys.Rev.C021301, Chu2010Phys.Rev.C024320} predict also a highly depressed density profile for $^{46}$Ar without the tensor force. On the other hand, the pairing correlation effect could significantly quench the bubble-like structure. In Ref. \cite{Nakada2013Phys.Rev.C067305}, with semirealistic $NN$ interaction, the proton bubble-like structure is unlikely to exist in the argon isotopes due to the strong pairing effects. Different from $^{46}$Ar, the emergence of the proton bubble-like structure in $^{34}$Si may be owing to the large gap between the proton $\pi2s_{1/2}$ and $\pi1d_{5/2}$ states, namely the $Z=14$ proton shell \cite{Piekarewicz2007J.Phys.G467}. However, as predicted by the particle-number and angular-momentum projected generator coordinate method (GCM) based on the mean-field approaches, the dynamical correlation might strongly quench the bubble-like structures in both $^{34}$Si \cite{Yao2012Phys.Rev.C014310, Yao2013Phys.Lett.B459} and $^{46}$Ar \cite{Wu2014Phys.Rev.C017304}. Therefore it is still an open question whether the proton bubble-like structure exists in $^{46}$Ar or $^{34}$Si. Besides these two, the central charge density of $^{44}$S is also predicted to be depressed distinctly in Ref. \cite{Chu2010Phys.Rev.C024320}, thus it is worthwhile to study the systematics of the charge density profiles along the $N = 28$ chains.

As addressed above, the emergence of the proton bubble-like structure in the nuclei around $^{46}$Ar is not only tightly related to the proton configurations near the Fermi surface, namely the position of the proton orbits $\pi2s$ and $\pi1d$ and the gaps between, but also to the effects of the pairing and dynamical correlations. In recent years, as the natural extension of the density-dependent relativistic Hartree-Fock (DDRHF) theory \cite{Long2006Phys.Lett.B640, Long2007Phys.Rev.C034314}, the relativistic Hartree-Fock-Bogoliubov (RHFB) theory with density-dependent meson-nucleon couplings was established to describe the weakly bound exotic nuclei \cite{Long2010Phys.Rev.C024308}. With the presence of the Fock terms in the mean-field channel, especially the inclusion of $\rho$-tensor couplings, substantial improvements have been achieved in the self-consistent descriptions of the nuclear shell structure and the evolution of the ordinary and superheavy nuclei \cite{Long2007Phys.Rev.C034314, Long2009Phys.Lett.B428,Li2014Phys.Lett.B169}, the nuclear novel phenomena including the halo structures in cerium and carbon isotopes \cite{Long2010Phys.Rev.C031302, Lu2013Phys.Rev.C034311}, and the nuclear spin-isospin excitation modes \cite{Liang2008Phys.Rev.Lett.122502, Liang2012PhysRevC.85.064302, Niu2013Phys.Lett.B723}.

Inspired by the above mentioned advantages, in this work we take the argon isotopes and $N=28$ isotones as the candidates to explore the occurrence of the proton bubble-like structure within the RHFB theory, as compared to the relativistic Hartree-Bogoliubov (RHB) calculations \cite{Niksic2011Prog.Part.Nucl.Phys66, Meng1998Nucl.Phys.A3, Vretenar2005Phys.Rep.101, Meng2006Prog.Part.Nucl.Phys.470}. Based on the calculations with both RHFB and RHB theories, the bulk properties, the charge-density profiles, and the proton single-particle levels of the selected nuclei will be analyzed in details. The contents are organized as follows. In Sec. \ref{Sec:method}, we briefly recall the general formalism of the RHFB theory and the charge-density profile. In Sec. \ref{Sec:results} are given the detailed discussions on the emergence of the proton bubble-like structures in the selected isotopes and isotones, including the model-reliability test, the charge-density profiles, the proton single-particle spectra, and the relevant pseudo-spin symmetry. Finally, a brief summary is drawn in Sec. \ref{Sec:summary}.

\section{The method}\label{Sec:method}

Under the framework of the DDRHF theory  and its extension --- the RHFB theory, the nucleons in finite nuclei are treated as point-like particles moving independently in the mean field provided by the others, i.e., the mean-field approach. Consistent with the meson-exchange picture of nuclear force, the model Lagrangian as the theoretical starting point is then constructed by including the degrees of freedom associated with the nucleon ($\psi$), the isoscalar $\sigma$- and $\omega$-mesons, the isovector $\rho$- and $\pi$-mesons, and the photon (A) \cite{Bouyssy1987Phys.Rev.C380, Long2006Phys.Lett.B640, Long2007Phys.Rev.C034314}. At the level of mean-field approach, the isoscalar $\sigma$- and $\omega$-meson fields dominate the nuclear attractive and repulsive interactions, respectively, and the isovector $\rho$- and $\pi$-meson degrees of freedom are responsible for the isospin-related aspects of nuclear force, and the photon field stands for the electro-magnetic interactions between protons. In fact, not only the isovector ones, the Fock diagrams of the isoscalar $\sigma$- and $\omega$-couplings also carry the isovector nature of nuclear force, showing substantial contributions to the symmetry energy \cite{Sun2008Phys.Rev.C065805, Long2012Phys.Rev.C025806}.

Following the standard variational procedure, one can derive the field equations of nucleon, mesons and photon from the Lagrangian, i.e., the Dirac, Klein-Gordon and Proca equations, respectively. Meanwhile, the continuity equation, leading to the energy-momentum conservation, can be also obtained, from which is derived the Hamiltonian of the system. In this work, which deals with the nuclear ground states, the time-component of the four-momentum carried by the mesons are dropped, which amounts to neglecting the retardation effects in the Fock terms \cite{Bouyssy1987Phys.Rev.C380}. Substituting the field equations of mesons and photon, the Hamiltonian can be expressed as the functional of nucleon field. In terms of the creation and annihilation operators $(c_\alpha^\dag,c_\alpha)$ defined by the stationary solutions of the Dirac equation, the Hamiltonian operator can be further quantized as
\begin{equation}\label{Hamiltonian}
  H=\sum_{\alpha\beta}c_\alpha^\dag c_\beta T_{\alpha\beta} + \frac{1}{2}\sum_{\alpha\alpha'\beta\beta'}c_\alpha^\dag c_\beta^\dag c_{\beta'}c_{\alpha'}\sum_\phi V_{\alpha\beta\alpha'\beta'}^\phi,
\end{equation}
where $T_{\alpha\beta}$ represents the kinetic energy term, and the two-body potential energy terms $V_{\alpha\beta\alpha^\prime\beta^\prime}^\phi$ correspond to the meson- (or photon-) nucleon couplings denoted by $\phi$,
\begin{align}
  T_{\alpha\beta}=&\int{}d\pmb r\bar\psi_\alpha(\pmb r)(-i\pmb{\gamma}\cdot\pmb{\nabla}+M)\psi_\beta(\pmb r),\\
  V_{\alpha\beta\alpha^\prime\beta^\prime}^\phi=&\int{}d\pmb rd\pmb r^\prime\bar\psi_\alpha(\pmb r)\bar\psi_\beta(\pmb r^\prime)\Gamma_\phi(\pmb r,\pmb r^\prime) \nonumber\\
  &\hspace{6em}\times{ }D_\phi(\pmb r,\pmb r^\prime)\psi_{\beta^\prime}(\pmb r^\prime)\psi_{\alpha^\prime}(\pmb r).
\end{align}
In the above expressions, $\psi_\alpha$ stands for the {Dirac spinor of nucleon}, $M$ is the nucleon mass, $\Gamma_\phi(\pmb r , \pmb r ^\prime)$ corresponds to the interaction matrices of various meson-nucleon coupling channels, i.e., the $\sigma$-scalar, $\omega$-vector, $\rho$-vector, $\rho$-tensor, $\rho$-vector-tensor, $\pi$-pseudo-vector, and photon-vector couplings, and $D_\phi(\pmb r , \pmb r ^\prime)$ represents the meson (photon) propagator.

In the limit of mean-field approach, the contributions from the negative energy states are generally neglected, namely the no-sea approximation. The nuclear energy functional $E$ is then determined as the expectation of the quantized Hamiltonian (\ref{Hamiltonian})  with respect to the Hartree-Fock ground state $|\Phi_0\rangle$,
\begin{equation}\label{Energy-functional}
  E=\langle\Phi_0|H|\Phi_0\rangle,\qquad\qquad|\Phi_0\rangle=\prod_{i}c_i^\dag|0\rangle,
\end{equation}
in which the index $i$ denotes the states corresponding to positive energy and $|0\rangle$ is the vacuum state. In contrast to the RHB theory, the RHFB theory includes both the Hartree and Fock mean fields which correspond to the direct and exchange terms in the expectation of the two-body potential $V^\phi$ with respect to the ground state $\left|\Phi_0\right>$, respectively.

For spherical nuclei, the relativistic Hartree-Fock (RHF) equation can be derived from the variation of the energy functional $E$ (\ref{Energy-functional}) with respect to the Dirac spinor $\psi(\pmb r )$ as \cite{Long2010Phys.Rev.C024308}:
\begin{equation}
  \int{}d\pmb r ^\prime{}h(\pmb r ,\pmb r ^\prime)\psi(\pmb r ^\prime)=\epsilon\psi(\pmb r),
\end{equation}
where $\epsilon$ is the single-particle energy and the single-particle Hamiltonian $h(\pmb r, \pmb r ^\prime)$ consists of three terms: the kinetic energy part $h^{\rm kin}$, the local and nonlocal potentials, i.e., $h^D$ and $h^E$, respectively,
\begin{subequations}
\begin{align}
  h^{\rm kin}(\pmb r ,\pmb r ^\prime)=&(\pmb\alpha\cdot\pmb{p}+\beta M)\delta(\pmb r -\pmb r ^\prime),\\
  h^{D}(\pmb r ,\pmb r ^\prime)=&[\Sigma_T(\pmb r )\gamma_5+\Sigma_0(\pmb r )+\gamma^0\Sigma_S(\pmb r )]\delta(\pmb r -\pmb r ^\prime),\\
  h^{E}(\pmb r ,\pmb r ^\prime)=&\left(\begin{array}{cc}
  Y_G(\pmb r ,\pmb r ^\prime)&Y_F(\pmb r ,\pmb r ^\prime)\\[0.5em]
  X_G(\pmb r ,\pmb r ^\prime)&X_F(\pmb r ,\pmb r ^\prime)
  \end{array}\right).
\end{align}
\end{subequations}
In $h^D$, the local potentials $\Sigma_S$, $\Sigma_0$ and $\Sigma_T$ include the Hartree mean fields and the rearrangement term, and the nonlocal ones $Y_G$, $Y_F$, $X_G$ and $X_F$ in $h^E$ correspond to the Fock mean fields \cite{Long2010Phys.Rev.C024308}.

For the open shell nuclei, one has to take the pairing correlations into account. In order to provide a reliable description, we incorporate the Bogoliubov scheme with the DDRHF theory to deal with the pairing effects, leading to the RHFB theory \cite{Long2010Phys.Rev.C024308}. Following the standard procedure of the Bogoliubov transformation \cite{Gorkov1959Sov.Phys.JETP1364, Kucharek1991Z.Phys.A23}, the RHFB equation can be derived as:
\begin{align}\label{RHFB}
&\int d\pmb r' \begin{pmatrix} h(\pmb r, \pmb r') & \Delta (\pmb r, \pmb r')\\[0.5em] -\Delta(\pmb r, \pmb r') & h(\pmb r, \pmb r')\end{pmatrix}\begin{pmatrix}\psi_U(\pmb r') \\[0.5em] \psi_V(\pmb r')\end{pmatrix} \nonumber\\
&\hspace{8em}= \begin{pmatrix}\lambda+E_q &0\\[0.5em] 0 &\lambda-E_q\end{pmatrix} \begin{pmatrix}\psi_U(\pmb r) \\[0.5em] \psi_V(\pmb r)\end{pmatrix}
\end{align}
where $\psi_U$ and $\psi_V$ represent the quasi-particle spinors, $E_q$ is the single quasi-particle energy, the chemical potential $\lambda$ is introduced to preserve the particle number on the average, and the pairing potential $\Delta(\pmb r , \pmb r ^\prime)$ reads as:
\begin{equation}
  \Delta_{\alpha}(\pmb r ,\pmb r ^\prime)=-\frac{1}{2}\sum_{\beta}V_{\alpha\beta}^{pp}(\pmb r ,\pmb r ^\prime)\kappa_{\beta}(\pmb r ,\pmb r ^\prime),
\end{equation}
with the pairing tensor $\kappa$,
\begin{equation}
  \kappa_\alpha(\pmb r ,\pmb r ^\prime)=\psi_{V_\alpha}(\pmb r )^\ast\psi_{U_\alpha}(\pmb r ^\prime).
\end{equation}
In the particle-particle ($pp$) channel, we utilize the finite-range Gogny force D1S \cite{Berger1984Nucl.Phys.A23} as the effective pairing force. Aiming at the nuclei around $^{46}$Ar, the original Gogny force D1S can provide appropriate descriptions of the pairing effects as demonstrated in Ref. \cite{Wang2013Phys.Rev.C054331}.  Notice that the RHFB equation (\ref{RHFB}) is a coupled integro-differential equation and is hard to be solved in coordinate space. In order to provide an appropriate description of the asymptotic behaviors of density profile, we expand the quasi-particle spinors $\psi_U$ and $\psi_V$ on the Dirac Woods-Saxon (DWS) basis \cite{Zhou2003Phys.Rev.C034323}, and the basis parameters, namely the spherical box size $R_{\max}$ and the numbers of positive and negative energy states (resp. $N_F$ and $N_D$), are taken as $R_{\max} = 28$ fm, $N_F = 44$, $N_D = 12$.

In this work, the charge density is determined from the proton-density profile by incorporating the corrections of the center-of-mass motion and finite nucleon size. The first correction is done by using the proton density in the center-of-mass reference frame, i.e., $\rho_{\rm c.m.}$ which is related to the Hartree-Fock (HF) proton density through,
\begin{equation}
  \rho_{\rm HF}(r) =\frac{4}{B^3\pi^{\frac{1}{2}}}\int e^{-r^{\prime2}/B^2}\rho_{\rm c.m.}(|r-r^\prime|)dr^\prime,
\end{equation}
where $B^{-2}= 2\big<\pmb P_{\rm c.m.}^2\big>/(3\hbar^2)$ and $\pmb P_{\rm c.m.}$ is the center-of-mass momentum. The second correction is taken into account by doing the convolution of $\rho_{\rm c.m.}$ with a Gaussian representing the form factor,
\begin{equation}
  \rho_{\rm ch}(r)=\frac{1}{2\pi^2r}\int_0^\infty k\sin{(kr)}\bar{\rho}_{\rm HF}(k)\exp\left[\frac{1}{4}k^2(B^2-a^2)\right]dk,
\end{equation}
where $\bar{\rho}_{\rm HF}(k)$ is the Fourier transform of the HF proton density and $a^2=2/3 \big(0.862^2 - 0.336^2 N/Z\big)$ accounts for the finite nucleon size \cite{Sugahara1994Nucl.Phys.A557}. Denoting $\lambda^2=1/(a^2-B^2)$, the charge-density distribution $\rho_{ch}$ is finally derived as,
\begin{equation}
  \rho_{\rm ch}(r)=\frac{\lambda}{\sqrt{\pi{}r^2}}\int{}r^\prime{}dr^\prime\rho_{\rm HF}(r^\prime) \left[e^{-\lambda^2(r-r^\prime)^2}-e^{-\lambda^2(r+r^\prime)^2}\right],
\end{equation}
where $\rho_{\rm HF}(r^\prime)$ corresponds to the proton density determined by the self-consistent calculations with the RH(F)B theories.


\section{RESULTS AND DICUSSIONS}\label{Sec:results}

In this paper, we focus on the occurrence of central depressions of the charge-density profiles of the nuclei around $^{46}$Ar, i.e., the proton bubble-like structure. The calculations are performed with the RHFB and RHB theories using the optimal effective interactions on the market, namely the RHF ones PKA1 \cite{Long2007Phys.Rev.C034314}, PKO1 \cite{Long2006Phys.Lett.B640} and PKO3 \cite{Long2008Europhys.Lett.12001}, and the RMF ones PKDD \cite{Long2004Phys.Rev.C034319} and DD-ME2 \cite{Lalazissis2005Phys.Rev.C024312}.

\begin{figure}[htbp]
  \includegraphics[width=0.45\textwidth]{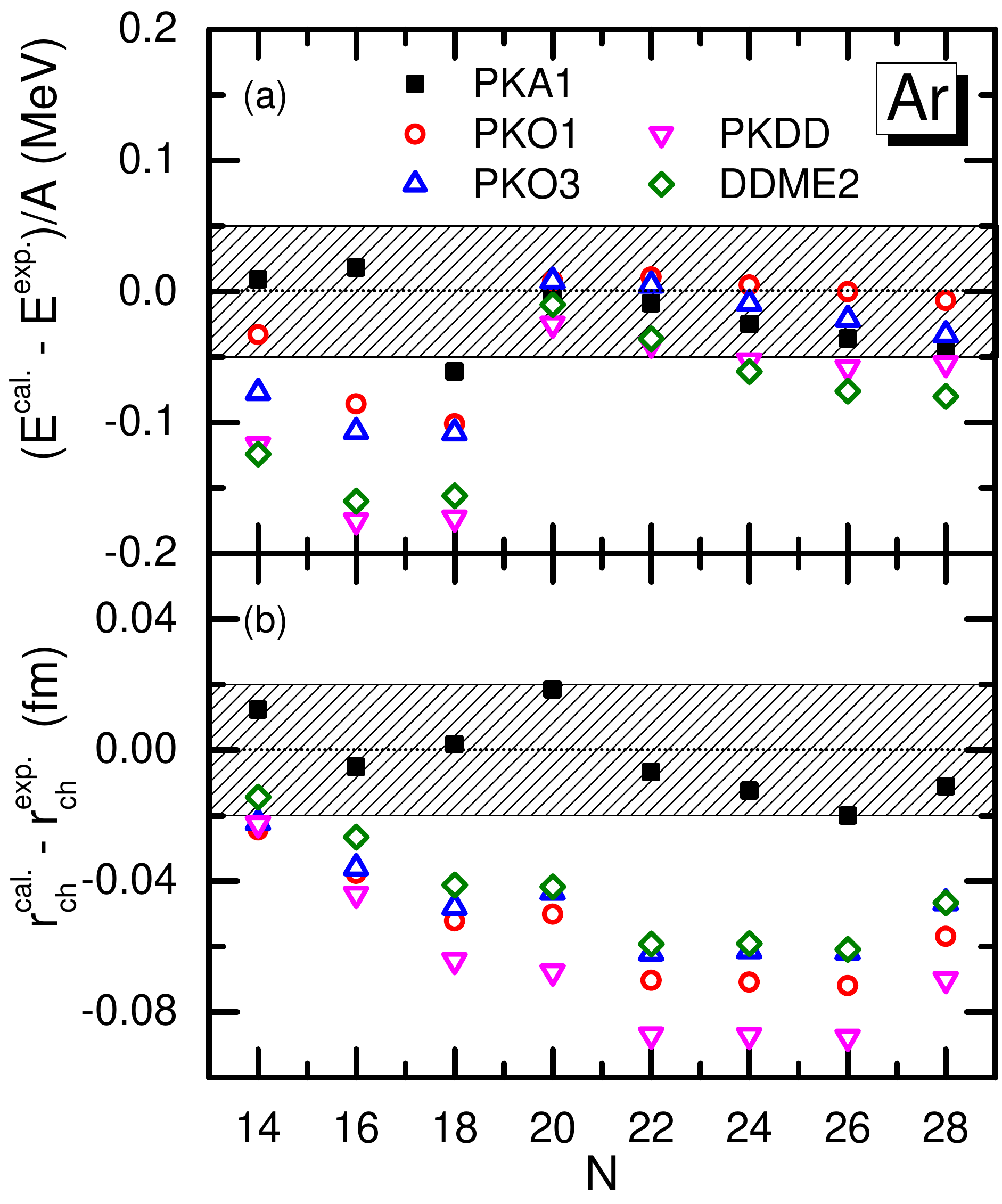}
  \caption{(Color online) Deviations of the calculated binding energies per particle $E/A$ (MeV) [plot (a)] and charge radii $r_{\rm ch}$ (fm) [plot (b)] from the experimental data \cite{Wang2012Chin.Phys.C1603, Angeli2013At.DataNucl.DataTables69} for argon isotopes. The theoretical results are extracted from the calculations of RHFB with PKA1, PKO1 and PKO3, and of RHB with PKDD and DD-ME2.}\label{fig:I}
\end{figure}

Taking argon isotopes as the representatives, we firstly test the model reliability in terms of the binding energies and the root-mean-sqaure (rms) charge radii, as referred to the experimental data \cite{Wang2012Chin.Phys.C1603, Angeli2013At.DataNucl.DataTables69}. Figures \ref{fig:I}(a) and \ref{fig:I}(b) display the deviations of the calculated binding energy per particle $E^{\rm cal.}/A$ and charge radii $r^{\rm cal.}_{\rm ch}$ from the data, respectively. It is found that both RHFB and RHB models provide appropriate agreements with the experimental data on the binding energies, whereas in the results of PKO1, PKO3, PKDD and DD-ME2 notable discrepancies appear on the neutron-deficient side. For the charge radius which contains both corrections of center-of-mass motion and finite nucleon size, the DDRHF functional PKA1 presents precise agreement with the data \cite{Angeli2013At.DataNucl.DataTables69}, and other selected models present relatively smaller values of $r_{\rm ch}$ than the data. Evidently, the RHFB theory with PKA1 provides the most reliable descriptions on the bulk properties of the argon isotopes, particularly the charge radii.

\begin{figure}[!htbp]
  \centering
  \includegraphics[width=0.47\textwidth]{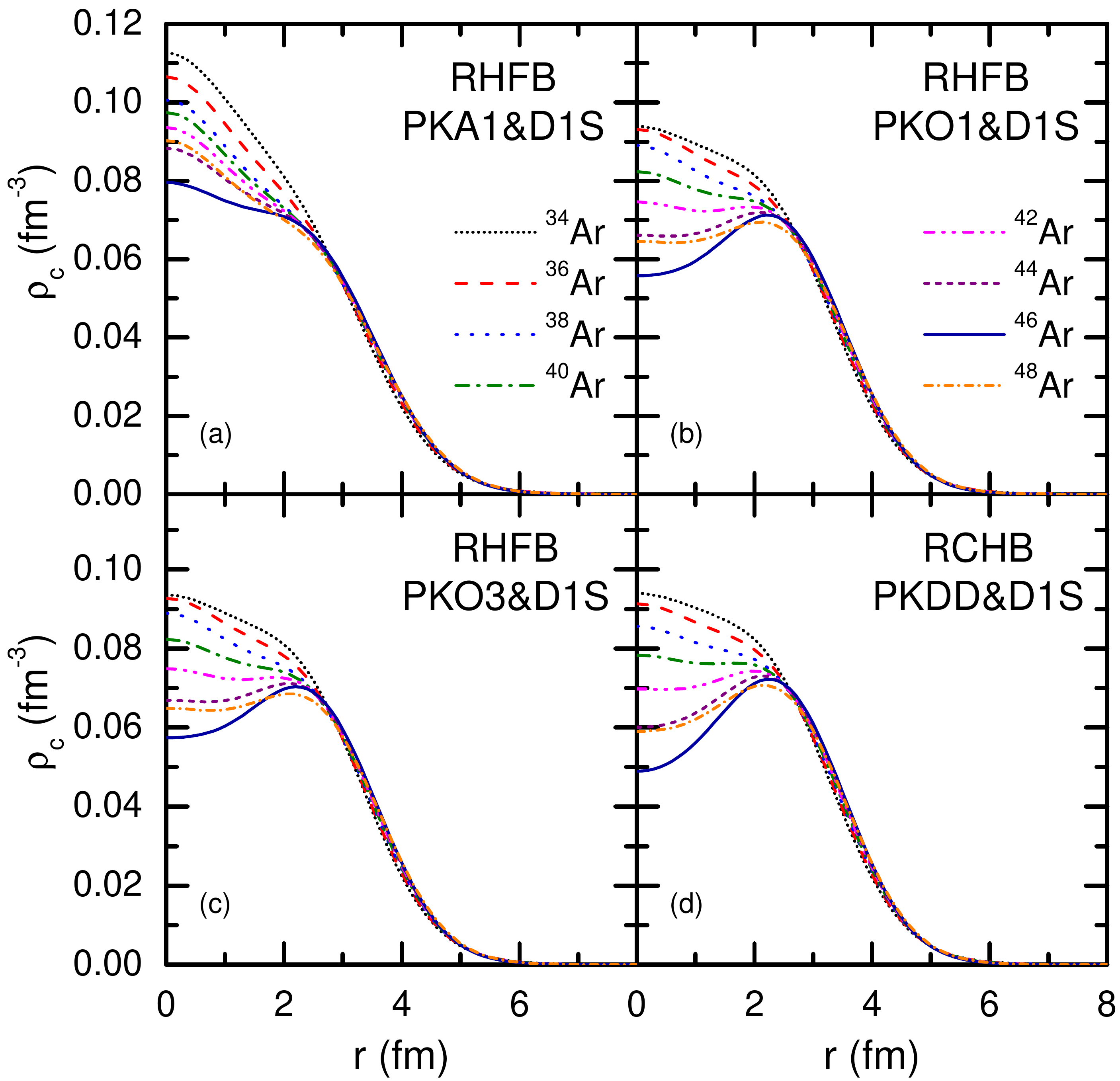}
  \caption{(Color online) The charge-density distributions of argon isotopes calculated by RHFB with PKA1, PKO1, PKO3 and RHB with PKDD.}\label{fig:II}
\end{figure}

As we mentioned in the introduction, $^{46}$Ar with two protons deficient from $^{48}$Ca was predicted as the candidate of a proton-bubble nucleus, which is characterized by the distinct central depression of charge-density distribution, if the inversion of the order of proton ($\pi$) orbits $\pi2s_{1/2}$ and $\pi1d_{3/2}$ occurred  \cite{Todd2004Phys.Rev.C021301, Khan2008Nucl.Phys.A37, Chu2010Phys.Rev.C024320, Wang2011Chin.Phys.Lett.102101, Wang2011Phys.Rev.C044333}. However, such inversion is essentially model dependent. To clarify the situation, we plot the charge-density profiles calculated by using the effective interactions PKA1, PKO1, PKO3 and PKDD respectively in Fig. \ref{fig:II} (a)-(d) for the argon isotopes. The results calculated with DD-ME2 is omitted, which shows similar systematics as PKO3 [see Fig. \ref{fig:II} (c)]. In Fig. \ref{fig:II}, it is clearly shown that the effective interactions PKO1, PKO3 and PKDD, which seem to predict $^{46}$Ar to have the bubble-like structure, present similar  charge-density profiles with distinct central depressions for the nuclei around $^{46}$Ar. On the contrary, the central depressions never appear in the charge-density profiles determined by PKA1 along the argon isotopic chain. Notice the fact that PKA1 provides the best agreement with the data of charge radii of argon isotopes among the selected models as shown in Fig. \ref{fig:I}(b). It seems that the occurrence of proton bubble-like structure is not favored in the argon isotopes.

\begin{figure}[!htbp]
  \centering
  \includegraphics[width=0.47\textwidth]{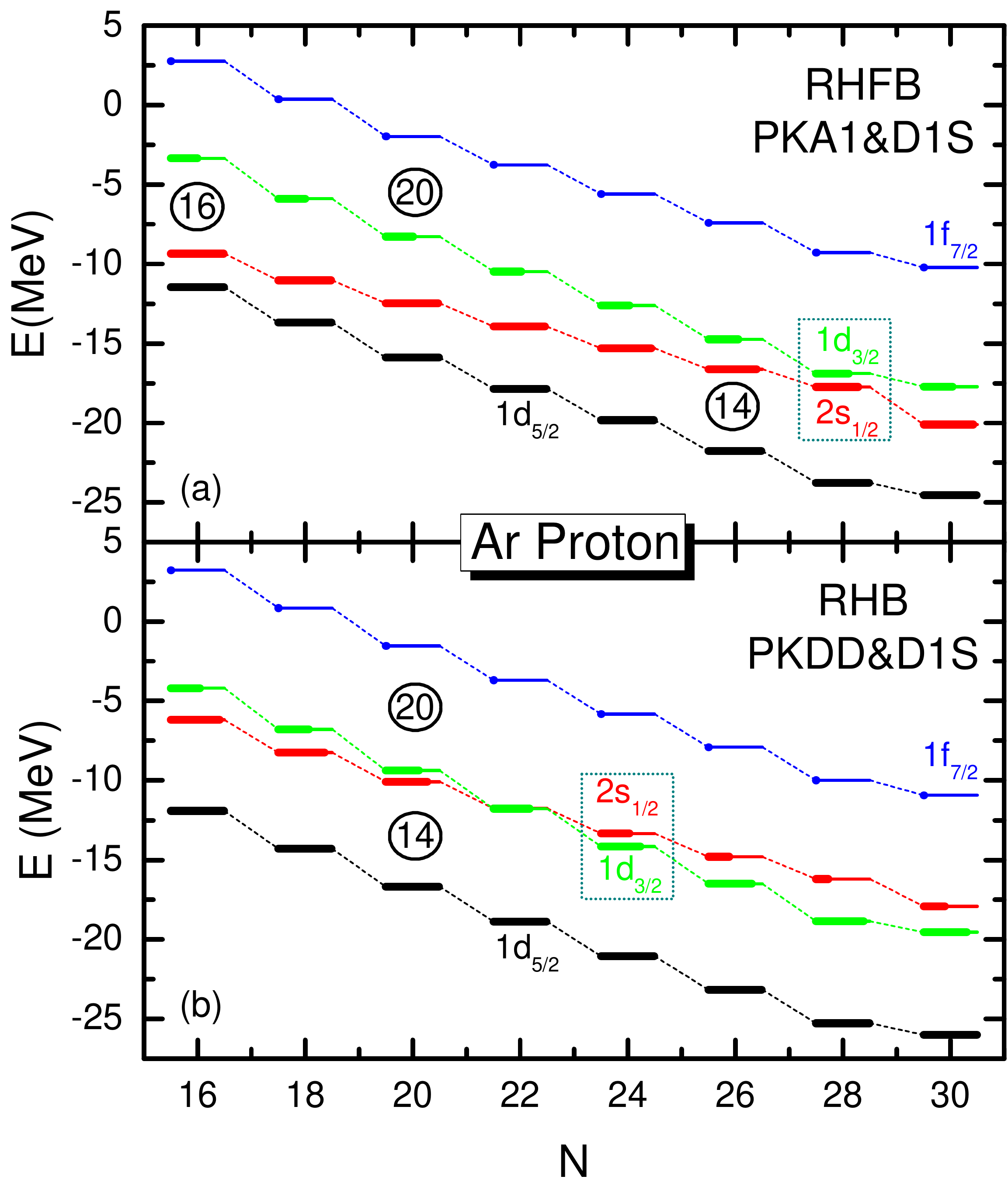}
  \caption{(Color online) Canonical proton single-particle energies for argon isotopes by RHFB with PKA1 and RHB with PKDD. The lengths of thick bars correspond to the occupation probabilities of proton orbits. }\label{fig:III}
\end{figure}

To understand the charge-density profiles of the argon isotopes, Fig. \ref{fig:III} shows the proton canonical single-particle spectra calculated by RHFB with PKA1 and RHB with PKDD, which present rather different charge distributions. In Fig. \ref{fig:III}, the lengths of the ultrathick bar denote the occupation probabilities of the orbits. As pointed out in Refs. \cite{Wang2011Chin.Phys.Lett.102101, Wang2011Phys.Rev.C044333, Nakada2013Phys.Rev.C067305}, the order of the proton ($\pi$) states $\pi1d_{3/2}$ and $\pi2s_{1/2}$ , as well as the gap between, is crucial for the occurrence of the bubble-like structure in $^{46}$Ar. Along the isotopic chain of argon, it is found from Fig. \ref{fig:III}(a) that PKA1 gives deeply bound and near fully occupied $\pi2s_{1/2}$ state, which does not support the formation of the bubble-like structure. While in Fig. \ref{fig:III}(b) the calculations with PKDD present an inversion on the order of the states $\pi1d_{3/2}$ and $\pi2s_{1/2}$ at $^{40}$Ar ($N=22$) and after that the proton state $\pi2s_{1/2}$ is less and less occupied, leading to the occurrence of the central depressions of charge density in $^{46}$Ar [see Fig. \ref{fig:II}(d)]. Concerning the shell closures $Z=14$ \cite{Piekarewicz2007J.Phys.G467} and $16$ \cite{Kanungo2002Phys.Lett.B58}, PKA1 presents distinct energy gap between $\pi1d_{3/2}$ and $\pi2s_{1/2}$ that gives the shell $Z=16$ at neutron-deficient side [see Fig. \ref{fig:III}(a)],  and approaching the neutron-rich side, this shell ($Z=16$) is strongly quenched and the one $Z=14$ emerges, leading to well preserved pseudo-spin symmetry, i.e., nearly degenerated proton orbits $\pi2s_{1/2}$ and $\pi1d_{3/2}$ in $^{46}$Ar ($N=28$). On the  contrary, the $Z=14$ shell is persistent well in the calculations with PKDD along the isotopic chain.

\begin{figure}[!htbp]
  \centering
  \includegraphics[width=0.47\textwidth]{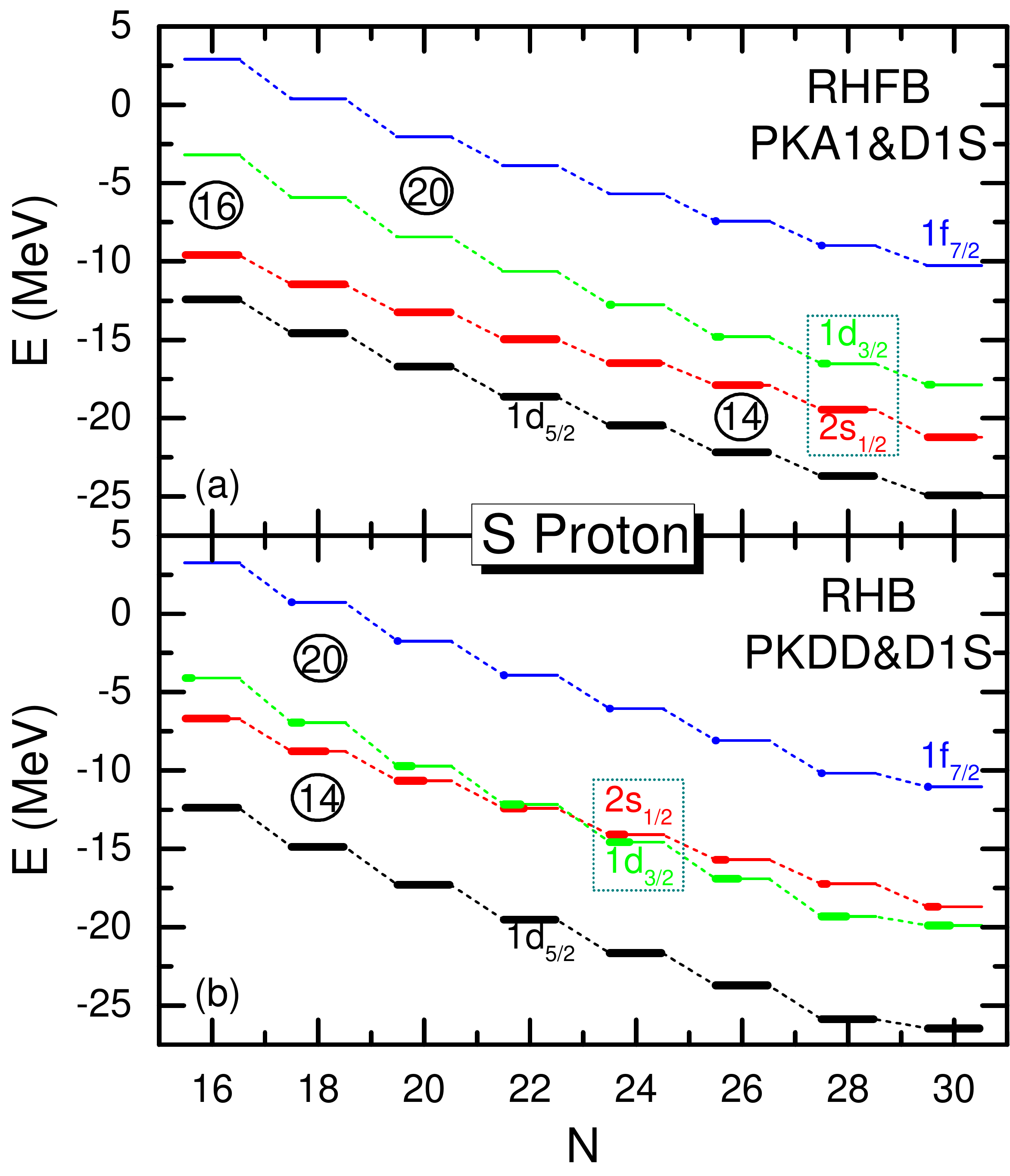}
  \caption{(Color online) Same as Fig. \ref{fig:III}, but for sulfur isotopes.}\label{fig:S-shell}
\end{figure}

It is worthwhile to mention that the analysis of beta-decay $Q$ values, single neutron separation energies, and the energies of the first excited state indicate the existence of the magic number $Z=16$ in neutron-rich regions of nuclear chart \cite{Kanungo2002Phys.Lett.B58}. To test the model reliability, Fig. \ref{fig:S-shell} shows the proton single-particle energies of the sulfur isotopes from $N=16$ to $30$, calculated by RHFB with PKA1 [Fig. \ref{fig:S-shell}(a)] and RHB with PKDD [Fig. \ref{fig:S-shell}(b)]. It is found that the calculations with PKA1 give consistent prediction on the emergence of the proton shell $Z=16$ with the analysis in Ref.  \cite{Kanungo2002Phys.Lett.B58}. While similar as argon isotopes, the RMF Lagrangian PKDD predicts only the proton shell $Z=14$ to occur in the proton spectra of sulfur isotopes. For the other selected Lagrangian, i.e., PKO1, PKO3 and DD-ME2, similar proton spectra are predicted as PKDD for both argon and sulfur isotopes.

As discussed in Refs. \cite{Todd2004Phys.Rev.C021301, Chu2010Phys.Rev.C024320, Khan2008Nucl.Phys.A37, Wang2011Chin.Phys.Lett.102101, Wang2011Phys.Rev.C044333}, the occurrence of the central depression of the charge-density profile in $^{46}$Ar is essentially related to the order of the proton states $\pi1d_{3/2}$ and $\pi2s_{1/2}$, as well as the energy gap between the states. Among the selected Lagrangians, PKO1, PKO3, PKDD and DD-ME2 seem to support the emergence of the bubble-like structure in the proton-density profile of $^{46}$Ar with similar mechanism as shown in Fig. \ref{fig:III}(b). However, the DDRHF Lagriangian PKA1, which has more complete meson-exchange diagram than the others, does not prefer the bubble-like structure to occur in the argon isotopes and such judgement is evidently supported by the fact that PKA1 presents better agreement on the charge radii of argon isotopes with the data \cite{Angeli2013At.DataNucl.DataTables69} than the others [see Fig. \ref{fig:I}(b)]. In particular, the proton shell $Z=16$ deduced from the systematical analysis of the experimental data (the beta-decay $Q$ values, single proton separation energies, and the energies of the first excited state) \cite{Kanungo2002Phys.Lett.B58} is properly reproduced only by PKA1 along the sulfur isotopic chain, from which the order of the states $\pi1d_{3/2}$ and $\pi2s_{1/2}$ is decided certainly with the experimental evidence. That is, the proton state $\pi2s_{1/2}$ must be deeper bound than $\pi1d_{3/2}$ to give the proton shell $Z=16$, and thus the occurrence of the bubble-like structure is prohibited.


\begin{table}
\caption{Binding energies per particle $E/A$ (in MeV) for $N=28$ isotones from $Z=12$ (Mg) to 20 (Ca) calculated by RH(F)B with PKA1, PKO1, PKO3, PKDD and DD-ME2. The experimental data are extracted from Ref. \cite{Wang2012Chin.Phys.C1603} and $\sigma$ stands for the root-mean square deviations from the data.}\label{tab:data1}
  \begin{ruledtabular}
    \begin{tabular}{ccccccc}
                 & Exp   & PKA1  & PKO1  & PKO3  &PKDD  & DD-ME2 \\      \hline
       $^{40}$Mg & 6.621 & 6.625 & 6.622 & 6.524 &6.458 & 6.456  \\
       $^{42}$Si & 7.416 & 7.405 & 7.392 & 7.317 &7.301 & 7.293  \\
       $^{44}$S  & 7.996 & 7.950 & 7.966 & 7.921 &7.896 & 7.877  \\
       $^{46}$Ar & 8.412 & 8.366 & 8.405 & 8.379 &8.357 & 8.332  \\
       $^{48}$Ca & 8.667 & 8.674 & 8.695 & 8.683 &8.665 & 8.644  \\
       $\sigma$  &       & 0.030 & 0.021 & 0.072 &0.103 & 0.113  \\
  \end{tabular}
  \end{ruledtabular}
\end{table}

\begin{figure}[!htbp]
  \centering
  \includegraphics[width=0.47\textwidth]{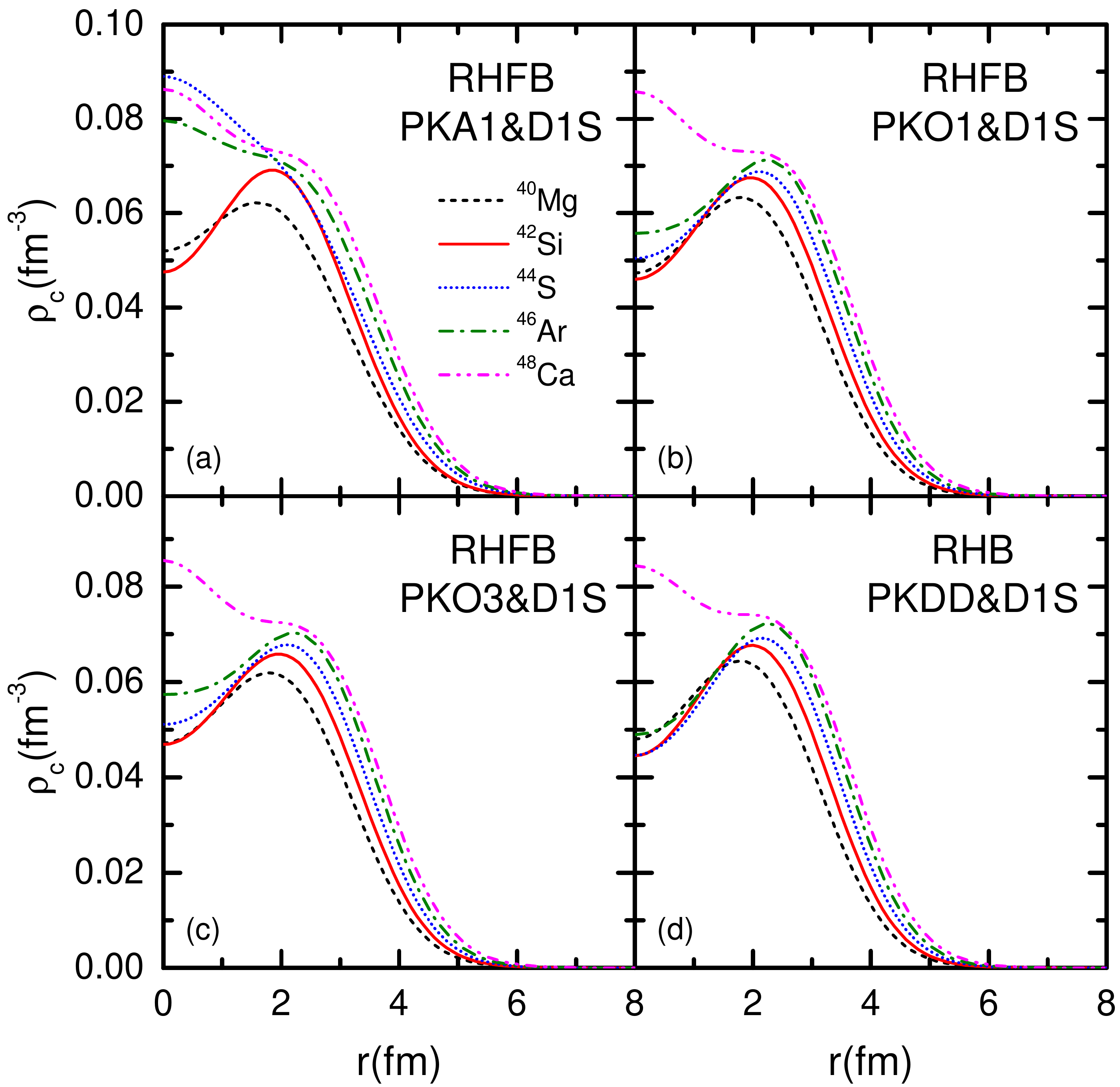}
  \caption{(Color online) The charge-density distributions of N=28 isotones calculated by RHFB with PKA1, PKO1, PKO3 and RHB with PKDD.}\label{fig:V}
\end{figure}

\begin{figure}[!htbp]
  \centering
  \includegraphics[width=0.47\textwidth]{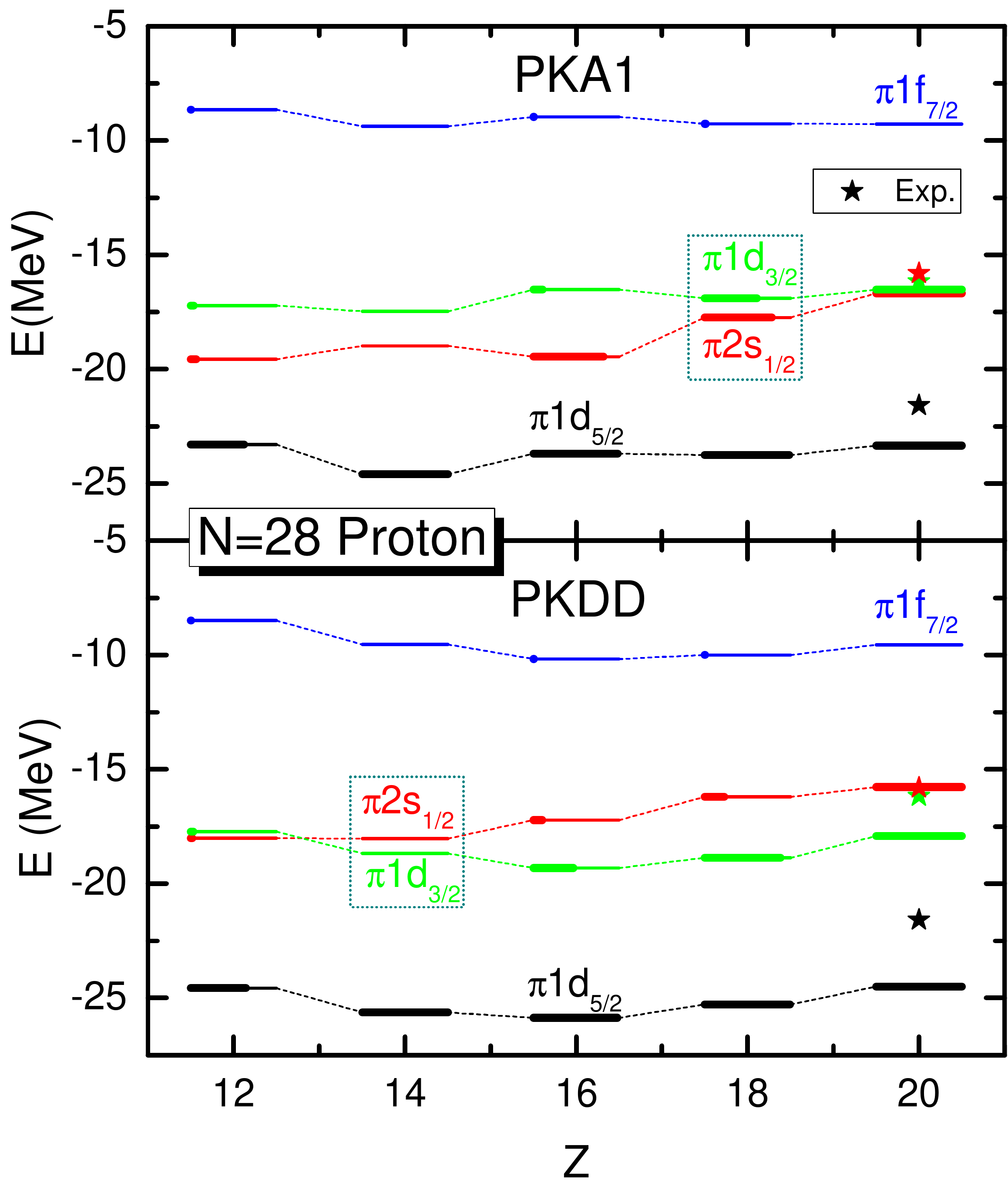}
  \caption{(Color online) Canonical proton single-particle energies along the $N=28$ isotonic chain calculated by RHFB with PKA1 and RHB with PKDD. The lengths of thick bars correspond with the occupation probabilities of the proton orbits and the filled stars denote the experimental data taken from Ref. \cite{Grawe2007Rep.Prog.Phys.1525}.}\label{fig:IV}
\end{figure}

To further probe the bubble-like structure, we extend the exploration from $^{46}$Ar along the isotonic chain of $N=28$, by taking the isotones from $^{40}$Mg to $^{48}$Ca. Table \ref{tab:data2} shows the binding energy per particle $E/A$ of the selected isotones calculated by RH(F)B with PKA1, PKO1, PKO3, PKDD and DD-ME2, in comparison with the data \cite{Wang2012Chin.Phys.C1603}. As seen from the root-mean square deviations in the last row, all the selected models present appropriate agreement with the data, and PKA1 and PKO1 reproduce the binding energies better than the others. Turning to the charge-density profiles in Figs. \ref{fig:V}(a)-(d) which respectively show the results calculated by RHFB with PKA1, PKO1\ and PKO3 and by RHB with PKDD, it is found that the calculations with PKO1, PKO3 and PKDD present similar charge-density distributions with each other, in which the bubble-like structures, i.e., the central depressions, are predicted for the selected isotones only except $^{48}$Ca [see Figs. \ref{fig:V}(b)-(d)]. While the central depressions exist only in the charge-density profiles of $^{40}$Mg and $^{42}$Si, as determined by the RHFB calculations with PKA1 [see Fig. \ref{fig:V}(a)]. Such distinct deviations between the models can be understood from the proton single-particle spectra shown in Fig. \ref{fig:IV}. It is seen that PKA1 predicts deeper bound $\pi2s_{1/2}$ state than $\pi1d_{3/2}$ for the selected isotones and the state $\pi2s_{1/2}$ is gradually occupied by the valence protons since $^{44}$S ($Z=16$). As a result, the formation of the bubble-like structure is blocked in the isotones from $^{44}$S to $^{48}$Ca. Different from PKA1, the calculations with PKDD present an inversion on the order of the states $\pi2s_{1/2}$ and $\pi1d_{3/2}$ from $Z=12$ ($^{40}$Mg) to $14$ ($^{42}$Si) and then the valence protons are filled mainly in the orbit $\pi1d_{3/2}$, leading to the distinct central depressions in the charge-density profiles from $^{44}$S to $^{46}$Ar. Similar systematics are also found in the proton spectra determined by PKO1, PKO3 and DD-ME2 as by PKDD.

In fact, as shown in Figs. \ref{fig:II}-\ref{fig:III}, i.e., the charge-density profiles and the corresponding proton spectra along the argon isotopic chain, the occurrence of the bubble-like structures is tightly related not only to the order of the states $\pi2s_{1/2}$ and $\pi1d_{3/2}$ but also to the splitting between these two pseudo-spin partners. As shown in Fig. \ref{fig:III}(b), although the order of $\pi2s_{1/2}$ and $\pi1d_{3/2}$ is reversed at $N=22$, the emergence of the proton bubble-like structure in $^{40-44}$Ar is still not favored very much [see Figs. \ref{fig:II}(b)-(d)] because of the fairly large occupations in $\pi2s_{1/2}$ induced by the pairing correlations, which is essentially influenced by the energy gap between the states. For the $N=28$ isotones, if referring to the experimental data \cite{Grawe2007Rep.Prog.Phys.1525} denoted by stars in Fig. \ref{fig:IV}, the pseudo-spin symmetry related to the pseudo-spin partners $\pi2s_{1/2}$ and $\pi1d_{3/2}$ in $^{48}$Ca is properly reproduced by PKA1, whereas the calculations with the others show distinct discrepancy from the data, e.g., PKDD presents notable splitting between the states $\pi2s_{1/2}$ and $\pi1d_{3/2}$ as shown in Fig. \ref{fig:IV}(b).

In order to better understand the deviations between the models in predicting the occurrence of the bubble-like structure, it is worthwhile to check the pseudo-spin symmetry in the relevant nuclei. From the Dirac equation, the single-particle energy of a state $\alpha$ can be expressed as,
\begin{equation}\label{Eq:single}
E_\alpha=E_{k, \alpha}+E_{\sigma, \alpha}+E_{\omega,\alpha}+E_{\rho,\alpha}+E_{\pi,\alpha}+E_{A,a}+E_{R,\alpha},
\end{equation}
where $E_{k,\alpha}$, $E_{\phi,\alpha}$ $(\phi=\sigma,\omega,\rho,\pi,A)$ and $E_{R,\alpha}$ denote the contribution of the kinetic energy, potential energy and rearrangement terms, respectively. According to Eq. (\ref{Eq:single}) and using the canonical wave functions determined from the RH(F)B calculations, the contribution to the splittings of the pseudo-spin partners $\pi2s_{1/2}$ and $\pi1d_{3/2}$ are determined by the selected Lagrangians for $^{48}$Ca. The results are shown in Table \ref{tab:data2}, including the experimental values of the average energy $\bar E =  \big(2E_{\pi2s_{1/2}} + 4E_{\pi1d_{3/2}}\big)/6 $ and the splitting $\Delta E$ as a reference. Identical with the proton spectra shown in Fig. \ref{fig:IV}, only PKA1 properly reproduce the splittings between the pseudo-spin partners $\pi2s_{1/2}$ and $\pi1d_{3/2}$. Specifically, the term $\Delta E_{\sigma+\omega}$ extracted from the calculations with PKA1, namely the balance between the strong $\sigma$- and $\omega$-meson fields, plays an important role in reducing the pseudo-spin orbital splitting, which indicates that PKA1 presents a difference balance between the nuclear attraction and repulsion from the others \cite{Long2007Phys.Rev.C034314}. Similar systematics are also found in the detailed contributions of the splittings of the partners $\pi2s_{1/2}$ and $\pi1d_{3/2}$ in $^{46}$Ar.

\renewcommand{\arraystretch}{1.1}
\begin{table}
\caption{Average energy $\bar E = \big(2E_{\pi2s_{1/2}} + 4E_{\pi1d_{3/2}}\big)/6$ (in MeV) and contributions (in MeV) to the splittings $\Delta E$ of the pseudo-spin partner states $\pi2s_{1/2}$ and $\pi1d_{3/2}$ in $^{48}$Ca, determined by the calculations with PKA1, PKO1, PKO3, PKDD and DD-ME2. The experimental values of $\bar E$ and $\Delta E$ are extracted from Ref. \cite{Grawe2007Rep.Prog.Phys.1525}. }\label{tab:data2}
  \begin{ruledtabular}
  \begin{tabular}{crrrrrr}
$^{48}$Ca                  & Exp.    &PKA1      &PKO1      &PKO3      &PKDD       &DD-ME2     \\ \hline
$\bar{E}$                  &$-$16.05 &$-$16.571 &$-$16.693 &$-$16.957 &$-$17.205  &$-$17.084  \\
$\Delta E$                 &0.36     &$-$0.170  &1.718     &1.716     &2.154      &1.516      \\ \hline
$\Delta E_k$               &$-$~~    &1.763     &1.152     &1.244     &1.229      &0.907      \\
$\Delta E_\rho$            &$-$~~    &1.366     &0.877     &0.640     &0.791      &0.489      \\
$\Delta E_\pi$             &$-$~~    &0.736     &0.569     &0.964     &$-$~~      &$-$~~      \\
$\Delta E_A$               &$-$~~    &0.149     &0.185     &0.197     &0.062      &$-$0.048   \\
$\Delta E_R$               &$-$~~    &$-$1.350  &$-$1.015  &$-$0.971  &$-$0.908   &$-$0.775   \\
$\Delta E_{\sigma+\omega}$ &$-$~~    &$-$2.961  &$-$0.185  &$-$0.492  &0.855      &0.828      \\  \hline
$\Delta E_{\sigma}$        &$-$~~    &5.907     &9.656     &8.620     &24.838     &28.649     \\
$\Delta E_{\omega}$        &$-$~~    &$-$8.868  &$-$9.841  &$-$9.113  &$-$23.983  &$-$27.821  \\

  \end{tabular}
  \end{ruledtabular}
\end{table}

Combining the results of the argon isotopes and $N=28$ isotones, it can be concluded that the bubble-like structure of the charge-density profiles is predicted to occur in the $N=28$ isotones $^{40}$Mg and $^{42}$Si commonly by the selected RHF and RMF Lagrangians, due to the fact that in these two isotones both the proton states $\pi2s_{1/2}$ and $\pi1d_{3/2}$ are not occupied. However, for the popular candidate $^{46}$Ar, the RHF model PKA1 does not support the occurrence of the bubble-like structure in the charge-density profiles, and evidently it can provide better agreement with the data of binding energies and charge radii of the argon isotopes than the other RHF and RMF models (see Fig. \ref{fig:I}), as well as the shell structures $Z=14$ and $16$ nearby [see Fig. \ref{fig:S-shell}(a)]. In addition, if starting from $^{48}$Ca in which the pseudo-spin symmetry related to the partner states $\pi2s_{1/2}$ and $\pi1d_{3/2}$ is demonstrated to be conserved experimentally, the neighbored $^{46}$Ar is expected to have nearly degenerated pseudo-spin doublet  $\pi2s_{1/2}$ and $\pi1d_{3/2}$ and consequently the occurrence of the bubble-like structure will be blocked by the pairing effects which lead to the spreading of the valence protons over these two states. Among the selected models, only the RHF model PKA1 presents consistent prediction on the conservation of pseudo-spin symmetry in $^{48}$Ca and $^{46}$Ar. On the other hand, the occurrence of the bubble-like structure is also tightly related to the order of the states $\pi2s_{1/2}$ and $\pi1d_{3/2}$. From Fig. \ref{fig:S-shell}(b), some sulfur isotopes seem to have bubble-like structure, according to the proton configurations determined by the RMF model PKDD. While if referring to the existence of the $Z=16$ proton shell as indicated by the experimental analysis \cite{Kanungo2002Phys.Lett.B58}, the bubble-like structure will not allow to occur as well. Eventually, the nuclei $^{40}$Mg and $^{42}$Si are predicted to have the proton bubble-like structure not only from the existence of the distinct central depressions in the charge-densitity distributions, but also from the proton single-particle configurations. According to the half-life of $^{42}$Si and $^{40}$Mg respectively as 12.5ms and 170ns \cite{Brookhaven}, $^{42}$Si may be treated as a potential candidate of proton-bubble nucleus for experimentalists.

\section{SUMMARY}\label{Sec:summary}
In this work we have studied the charge-density profiles and the proton spectra of the argon isotopes and $N=28$ isotones with the relativistic Hartree-Fock-Bogliubov (RHFB) theory using the effective interactions PKA1, PKO1 and PKO3, and with the relativistic Hartree-Bogliubov (RHB) theory using PKDD and DD-ME2. It is found that both models can reproduce the binding energies and charge radii of the argon isotopes with certain quantitative precision. Specifically, the PKA1 effective interaction provides the best agreements with the data, particularly on the emergence of the proton shells $Z=14$ and $16$ nearby,  and therefore the RHFB+PKA1 model is supposed to be the most reliable one among the selected models. In the calculations with PKA1, the inversion of the proton orbits $\pi2s_{1/2}$ and $\pi1d_{3/2}$ is not found in the argon isotopes to support the occurrence of the proton bubble-like structure. Along the isotonic chain of $N=28$, fairly distinct central depressions are found in the charge-density profiles of $^{42}$Si and $^{40}$Mg from the calculations of all the selected models, which are mainly due to the fact that the proton orbits  $\pi2s_{1/2}$ and $\pi1d_{3/2}$ is not occupied by the valence protons, and  $^{42}$Si may be treated as a potential candidate of proton bubble nucleus with longer life time than $^{40}$Mg. In addition it has been noted that another anti-bubble effect, namely the dynamical correlation, would quench the bubble structure in the ground-state of $^{34}$Si strongly \cite{Yao2013Phys.Lett.B459, Yao2012Phys.Rev.C014310}. In present work, $^{42}$Si is assumed to be the spherical nucleus, and perspectively it is interesting to test the existence of the bubble structure in $^{42}$Si after taking the dynamical correlation into account.

\section*{ACKNOWLEDGMENTS}
This work is partly supported by the National Natural Science Foundation of China under Grant No. 11375076, and the Specialized Research Fund for the Doctoral Program of Higher Education under Grant No. 20130211110005.


%

\end{document}